# Measurement of Local Reactive and Resistive Photoresponse of a Superconducting Microwave Device

A. P. Zhuravel

*B. Verkin Institute for Low Temperature Physics & Engineering, NAS of Ukraine, 47 Lenin Avenue, Kharkov, 61103, Ukraine*

Steven M. Anlage

*Physics Department, Center for Superconductivity Research, University of Maryland, College Park, MD 20742-4111 USA*

A. V. Ustinov

*Physics Institute III, University of Erlangen-Nuremberg, Erwin-Rommel Str. 1, D-91058, Erlangen, Germany*

(Dated: December 22, 2005)

We propose and demonstrate a spatial partition method for the high-frequency photo-response of superconducting devices correlated with inductive and resistive changes in microwave impedance. Using a laser scanning microscope, we show that resistive losses are mainly produced by local defects at microstrip edges and by intergrain weak links in the high-temperature superconducting material. These defects initiate nonlinear high-frequency response due to overcritical current densities and entry of vortices.

High-$T_c$ superconducting (HTS) materials have been successfully applied in passive high-frequency (HF) devices such as filters and resonators. Compared to their normal-metal counterparts the important advantage of superconducting (SC) devices is their small HF absorption, providing ultrahigh ($>10^6$) quality factor, Q [1]. The power handling capability of HTS systems is limited by strong nonlinearity (NL) of the surface impedance in d-wave cuprates at high exciting HF field. This manifests itself as generation of spurious harmonics and intermodulation product distortion.

Despite the voluminous work on the nature of NL effects in HTS, the causes are not completely clear and remain under debate [2]. It is widely accepted that sources of this problem have both resistive and inductive origins, and arise locally as a result of high current densities $J_{RF}(x,y)$ nonuniformly distributed in the cross-section of SC films. The lower limit of NLs can be attributed to the Nonlinear Meissner Effect (NLME), parameterized as a HF current (field) dependent magnetic penetration depth [3, 4];

$$\lambda_{eff}(T, J_{RF}) \cong \lambda_{eff}(T) * \left[1 + \frac{1}{2}\left(\frac{J_{RF}(x,y)}{J_{NL}(x,y)}\right)^2\right]$$

Here $\lambda_{eff}(T)$ is the local value of the temperature dependent magnetic penetration depth and $J_{NL}$ is a material parameter that corresponds to the intrinsic depairing current density in HTS. Large current densities can be reached at the strip edge because of the non-uniform Meissner screening even at modest input power levels. For currents above a current scale on the order of $J_{NL}$, penetration of vortices, hysteresis, and thermal dissipation contribute to the NL of the HTS device. Inhomogeneities generate additional resistive NL sources, with smaller values of $J_{NL}$, reducing the power handling capability even further. The standard methods to asses NL essentially integrate the response over the whole device [5, 6] and cannot solve this problem because multiple NL sources are scattered inhomogeneously about the HTS film. Therefore it is highly desirable to develop new methods of local analysis capable not only to find NL sources, but also to correlate their location and effects with other properties of HTS materials, such as $J_{RF}(x,y)$ and structural perfection.

A laser scanning microscope (LSM) can measure optical and HF properties in HTS devices simultaneously and is ideally suited for studying the above-stated problems. The LSM uses the principle of raster scanning the surface of a sample by a sharply focused laser beam (probe). The *(i)* reflected laser power is detected as a function of probe coordinate x,y to image optically distinguishable defects of its geometry and structure. The *(ii)* absorbed part of laser energy heats the sample on the thermal healing length scale, $l_T$ for bolometric probing of all the thermosensitive properties of the device, including the microwave transmittance $S_{21}(f)$. In this case, the thermally induced changes of $S_{21}(f)$ in the probe are understood as LSM photoresponse (PR) that can be expressed as:

$$PR \propto \frac{\partial \|S_{12}(f)\|^2}{\partial T} \delta T .$$

In particular, as has been postulated earlier, modulation of kinetic inductance by a thermal probe allows one to measure a quantity proportional to $A\lambda_{eff}^2(x,y)J_{RF}^2(x,y)\delta\lambda_{eff}$, and consequently can be used for extraction of $J_{RF}^2(x,y)$ maps [7]. Here $\delta\lambda_{eff}$ is the photo-induced change in $\lambda_{eff}$, and $A$ is the area of the thermal spot (related to $l_T$). However, as will be shown below, the local LSM bolometric PR has a more complex nature and contains a resistive component in addition to the inductive one. The amplitude of the (HF current driven) resistive LSM PR component is proportional to probe-induced dissipation in HTS and is therefore also proportional to $J_{RF}^2(x,y)$. In this letter, we propose and demonstrate a way to separate resistive and inductive components of PR based on a new procedure of collecting and processing LSM images.

The examined sample was a co-planar wave-guide (CPW) resonator designed for operation with fundamental resonant frequency of about $f_0$= 5.6 GHz with $Q$ = 600 at a temperature of 78 K. The HTS structure was patterned by wet lithography from a $YBa_2Cu_3O_{7-\delta}$ (YBCO) film with thickness $d$ = 240±28 nm laser ablated on a 500 μm $LaAlO_3$ ($\varepsilon_r$=24.2) substrate (LAO). The CPW of $w$=500 μm width, $L_0$=7.75 мм length and 650 μm gap from the ground plane was capacitively coupled to SMA connectors attached to a cooled Cu sample fixture. The measurements were performed in an optical cryostat stabilizing temperature of the sample between 77 K and $T_c$ = 86.4 K with an accuracy of 0.005 K.

The SC device was cooled below $T_c$ and the resonant mode was excited by an external synthesizer at a frequency f, producing the transmission spectrum $|S_{21}(f)|^2$ shown in Fig. 1. The surface of the CPW resonator was scanned by a laser beam (wavelength: 670 nanometers, power 1 mW) focused in an optical probe of 1.2 μm diameter. The intensity of the laser is square-wave modulated by an oscillator with frequency $f_M = \omega_M/2\pi$ = 100 kHz, raising the average temperature of the sample by $\Delta T$ = 0.2 K and creating a thermal probe (of $2l_T$ diameter) oscillating in the center with magnitude $(\Delta T/2) \cos(\omega_M t + \phi)$, where $\phi$ is the phase shift of the temperature oscillation relative to the square wave modulation [8]. A good acoustic match between the YBCO film and LAO substrate of the device at temperatures above 50 K leads to an $l_T$ that is defined mainly by the properties of LAO, producing a thermal spot with $l_T=(k/c\rho f_M)^{1/2}$ = 4.8 μm with thermal conductivity k = 9 W/m*K, a specific heat c = 0.58 J/g*K, and density $\rho$ = 6.57 g/cm$^3$. This estimate of $l_T$ is in good agreement with the measured value [9].

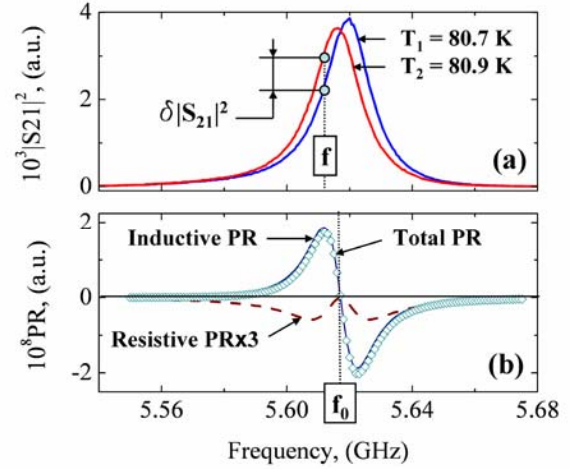

FIG. 1. (a) Temperature modification of the microwave transmittance $|S_{21}|^2(f)$ of a CPW resonator at $P_{IN}$=0 dBm and $\Delta T=T_2-T_1$=0.2 K, (b) difference (symbols) that is proportional to the sum of LSM PR, along with the inductive (solid line) and resistive (dashed line) components. The symmetry of the resistive component and asymmetry of the inductive component relative to $f_0$ is clearly visible.

The oscillating local heating of the HTS film causes the thermo-induced modulation of transmitted HF power $\delta P_{OUT}(f) \sim \delta\|S_{21}(f)\|^2$, which was detected by a spectrum analyzer and was used as a signal of the LSM PR, creating local contrast of LSM images. The typical ratio $\delta P_{OUT}/P_{OUT} \sim 10^{-6}$ depends on the ratio between the sizes of the probe and the sample. Since the variation in $\delta P_{OUT}$ is on the level of HF noise, increased sensitivity is achieved by using ac lock-in detection of the spectrum analyzer signal synchronized to $\omega_M$.

Figure 1(a) illustrates the method of extraction for the averaged properties from our data [8]. The bolometric LSM PR is modeled here with a shifting of HF transmittance $\delta\|S_{21}(f)\|^2$ of the CPW resonator by the LSM probe. In this case, the LSM PR is generated (at any fixed HF frequency f) by a jump between temperatures $T_1$ and $T_2$ corresponding to the Lorenz-like profiles of $S_{21}(f)$ characteristic of the unperturbed and perturbed resonator. The resulting (taking into account the size of the thermal probe) LSM PR is shown by symbols in Fig. 1(b). Curves $T_1$ and $T_2$ in Fig. 1(a) are the experimental $\|S_{21}(f)\|^2$ spectra of the resonator obtained at $P_{IN} = 0$ dBm and at temperatures 80.7 K and 80.9 K, accordingly. The total $\delta\|S_{21}(f)\|^2$ has contributions from the effects of HF resonant frequency tuning $\delta f_0$ and broadening of the spectrum $\Delta f_{3dB}$. Generally, $\delta f_0$ is associated with the change of kinetic inductance $L_{ki} = \dfrac{\mu_0}{2}\dfrac{l}{w_S}\lambda_{eff}(T)$ due to the thermal probe, while $\Delta f_{3dB} \sim \Delta(1/2Q)$ is directly related to photo-induced modulation of the inverse Q-factor due to an increase in local Ohmic dissipation. Their separate contributions to total LSM PR can be estimated [8] from the resonance line shape is represented by:

$$\|S_{21}(f)\|^2 = \frac{S_{21}(f=f_0)}{1+4Q^2(f/f_0-1)^2}.$$

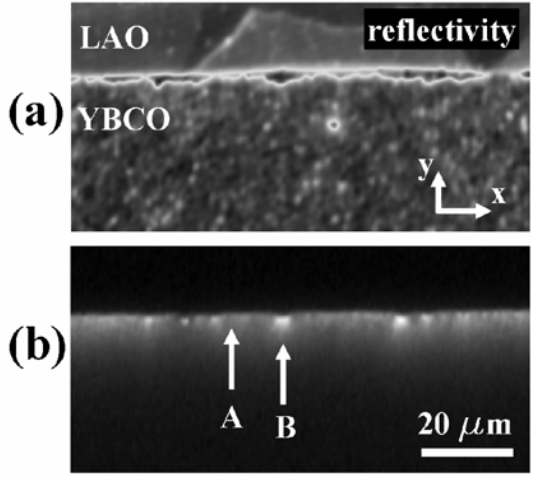

FIG. 2. (a) LSM reflectivity map showing the twinned structure of the LAO substrate that can lead to (b) inhomogeneous HF photoresponse in the YBCO film. Arrows indicate sections A and B corresponding to large difference in LSM PR selected for detailed analysis.

Considering, that both $Q$ and $f_0$ are temperature dependent and can be measured from a set of S-characteristics, inductive $PR_X$ and resistive $PR_R$ components of $\delta\|S_{21}(f)\|^2$ are obtained through partial derivatives:

$$PR_X \propto P_{IN} \frac{\partial \|S_{12}(f)\|^2}{\partial f_0} \delta f_0$$

and

$$PR_R \propto P_{IN} \frac{\partial \|S_{12}(f)\|^2}{\partial (1/2Q)} \delta(1/2Q).$$

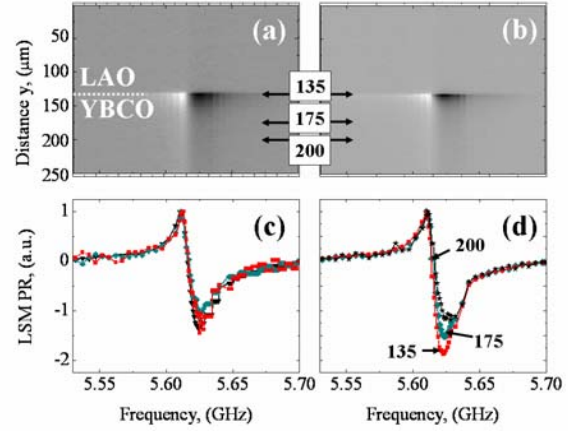

FIG. 3. Grey-scale representation of a frequency dependent LSM PR in sections (a) A and (b) B and the corresponding (c,d) amplitude profiles showing different decay of LSM PR in YBCO film due to the influence of twin-domains.

The extracted $PR_R$ and $PR_X$ are shown by dashed and solid lines, respectively, in Fig.1(b). Here the symbols show the difference between curves $T_2$ and $T_1$. At this temperature $PR_R$ is making less than a 15% contribution to the total LSM PR, and the averaged PR is generated mainly by its inductive component.

For comparison, the total local LSM PR has been measured at different areas of the CPW resonator under the same operating conditions. As an example, a 50x100 μm$^2$ scan is chosen in the vicinity of an edge of the YBCO center strip [see the reflective 100x50 μm$^2$ LSM image in Fig. 2(a)] close to the point of maximum $J_{RF}(x,y)$ in the standing wave pattern. The substrate clearly shows the patterns of twin domain blocks which modulate the LSM PR [see Fig. 2(b)] along the patterned edge of the YBCO film. Detailed analysis of LAO twinning on the total $J_{RF}(x,y)$ was done in an earlier paper [10]. Here we are interested in elucidating the microscopic origin of the LSM PR.

Figures 3(a) and (b) show frequency dependences of LSM PR in sections A and B, respectively, of Fig. 2(b) corresponding to the greatest contrast in $J_{RF}(x,y)$ inside the scanned area. A gray-scale representation is used for the PR(y, f) LSM images, where black areas correspond to (negative) minimum of LSM PR while the brightest areas correspond to (positive) maximum response, and gray corresponds to zero. The images are acquired by repeating a 250 μm y-line-scan at locations A and B at different HF frequency f, incremented in discrete steps of 0.1 MHz. The raster scan starts on LAO and crosses to YBCO at y=135 μm. It is clear from the marked horizontal sections of the LSM images in Fig. 3(a) and (b) that there is a different frequency dependence of the LSM PR at different points inside the HTS strip. As seen from Fig.3(c), the form of the normalized PR(f) in section

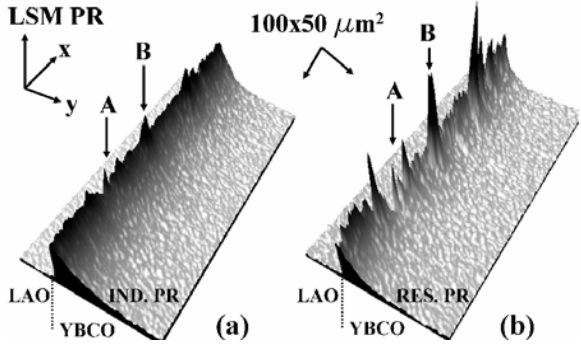

FIG. 4. 2D LSM images of (a) inductive and (b) resistive components of LSM PR, extracted by using new procedures described in text.

A practically remains invariant as a function of distance from the YBCO edge deep into the strip. At the same time, edge currents in section B give a sharp increase of negative LSM PR not far from the edge. The reason for this behavior is an increase of $PR_R$ induced by Cooper pair breaking and the generation of vortices where $J_{RF}(x,y) > J_C(x,y)$ in the "weak" section B.

One interesting feature is observed in the frequency dependence of the LSM PR [see Fig. 1(b)]. Its inductive component changes sign above and below $f_0$. In contrast, the resistive component is always negative and symmetrically positioned relative to $f_0$. This characteristic can be used for separation of the relative contribution to total PR as follows. Designate by $PR(f+)$ and $PR(f-)$ the LSM PR at equidistant frequencies $f+$ (above) and $f-$ (below) $f_0$, then

$$PR_R = \frac{|PR(f+) + PR(f-)|}{2} \quad (1)$$

gives the resistive component, while the procedure of subtraction

$$PR_X = \frac{|PR(f+) - PR(f-)|}{2} \quad (2)$$

gives the inductive contribution. Applying this procedure to the data in Fig. 3, the ratio $PR_X/PR_S=0.15$ is found for section A and is in good agreement with the bolometric CPW resonator volume averaged data presented in Fig.1. The edge HF currents in section B increase this ratio to 0.3 showing an almost two-fold increase of resistive losses due to the changes in edge current distribution caused by substrate twinning.

A similar procedure of component partition has been applied for improved 2D LSM imaging using Eqs. (1) and (2). Figure 4(a) shows the distribution $PR_X(x,y)$ along the edge of an element of the resonator from Fig. 2. The main structure of Fig. 4(a) is due to the rf current distribution $J_{RF}^2(x,y)$. The small spatial modulation of $PR_X$ seen here is related to local changes in $\lambda_{eff}(T,J)$ described above. Figure 4(b) shows the distribution of $PR_R(x,y)$ in the same area of LSM scanning. The overall average magnitude of $PR_R$ is about 2.67 times less than the $PR_X$ signal. However, the $PR_R$ response is highly localized and strongly fluctuating along the edge of the YBCO strip. Large peaks of $PR_R(x,y)$ on defects along the edge of the strip are the main candidates for local sources of NL in the resonator. This interpretation is consistent with the results of Booth *et al.* on YBCO where they attributed resistive NL to vortex entry at the edges of their CPW transmission lines [11].

In summary, we have developed a new LSM method for a spatial partition of the HF bolometric response of HTS devices correlated with inductive and resistive changes in microwave impedance. It is shown that resistive losses are mainly produced by local defects at microstrip edges and by intergrain weak links in the HTS material. These defects can initiate occurrence of NL HF response due to overcritical current densities and vortex entry.

We thank K. Harshavardhan of Neocera, Inc. for providing the CPW resonator samples. We acknowledge the support NSF/GOALI DMR-0201261, a NASU program on "nano-structures, materials and technologies", and German Science Foundation (DFG).